\documentclass[%
 reprint,
groupedaddress,
 amsmath,amssymb,
 aps,
prc,
]{revtex4-1}

\usepackage{graphicx}
\usepackage{dcolumn}
\usepackage{bm}
\usepackage{lineno}

\usepackage{booktabs}
\newcolumntype{C}[1]{>{\centering\arraybackslash}p{#1}}
\AtBeginDocument{
\heavyrulewidth=.08em
\lightrulewidth=.05em
\cmidrulewidth=.03em
\belowrulesep=.65ex
\belowbottomsep=0pt
\aboverulesep=.4ex
\abovetopsep=0pt
\cmidrulesep=\doublerulesep
\cmidrulekern=.5em
\defaultaddspace=.5em
}

\begin{document}

\preprint{APS/123-QED}


\title{Determination of Beta Decay Ground State Feeding of Nuclei of Importance for Reactor Applications}




\author{V. Guadilla}
\altaffiliation[Present address: ]{Faculty of Physics, University of Warsaw, 02-093 Warsaw, Poland}
\affiliation{%
 Instituto de F\'isica Corpuscular, CSIC-Universidad de Valencia, E-46071, Valencia, Spain
}
\author{J. L. Tain}%
\affiliation{%
 Instituto de F\'isica Corpuscular, CSIC-Universidad de Valencia, E-46071, Valencia, Spain
}
\author{A. Algora}%
\altaffiliation{%
 Institute of Nuclear Research of the Hungarian Academy of Sciences, Debrecen H-4026, Hungary
}
\affiliation{%
 Instituto de F\'isica Corpuscular, CSIC-Universidad de Valencia, E-46071, Valencia, Spain
}

\author{J. Agramunt}  
\affiliation{%
 Instituto de F\'isica Corpuscular, CSIC-Universidad de Valencia, E-46071, Valencia, Spain
}
\author{J. A. Briz}  
\affiliation{%
 Subatech, IMT-Atlantique, Universit\'e de Nantes, CNRS-IN2P3, F-44307, Nantes, France
}
\author{J. \"Ayst\"o}  
\affiliation{%
 University of Jyv\"askyl\"a, FIN-40014, Jyv\"askyl\"a, Finland
}
\author{A. Cucoanes}  
\affiliation{%
 Subatech, IMT-Atlantique, Universit\'e de Nantes, CNRS-IN2P3, F-44307, Nantes, France
}
\author{T. Eronen}  
\affiliation{%
 University of Jyv\"askyl\"a, FIN-40014, Jyv\"askyl\"a, Finland
}
\author{M. Estienne}  
\affiliation{%
 Subatech, IMT-Atlantique, Universit\'e de Nantes, CNRS-IN2P3, F-44307, Nantes, France
}
\author{M. Fallot}  
\affiliation{%
 Subatech, IMT-Atlantique, Universit\'e de Nantes, CNRS-IN2P3, F-44307, Nantes, France
}
\author{L. M. Fraile}  
\affiliation{%
Universidad Complutense, Grupo de F\'isica Nuclear, CEI Moncloa, E-28040, Madrid, Spain
}
\author{E. Ganio\u{g}lu}  
\affiliation{%
Department of Physics, Istanbul University, 34134, Istanbul, Turkey
}
\author{W. Gelletly}  
\affiliation{%
Department of Physics, University of Surrey, GU2 7XH, Guildford, UK
} 
\author{D. Gorelov}  
\author{J. Hakala} 
\author{A. Jokinen}
\affiliation{%
 University of Jyv\"askyl\"a, FIN-40014, Jyv\"askyl\"a, Finland
}
\author{D. Jordan} 
\affiliation{%
 Instituto de F\'isica Corpuscular, CSIC-Universidad de Valencia, E-46071, Valencia, Spain
}  
\author{A. Kankainen} 
\affiliation{%
 University of Jyv\"askyl\"a, FIN-40014, Jyv\"askyl\"a, Finland
} 
\author{V.S. Kolhinen} 
\altaffiliation[Present address: ]{%
 Cyclotron Institute, Texas A\&M University, College Station, Texas 77843, USA
}
\affiliation{%
 University of Jyv\"askyl\"a, FIN-40014, Jyv\"askyl\"a, Finland
}
\author{J. Koponen}  
\affiliation{%
 University of Jyv\"askyl\"a, FIN-40014, Jyv\"askyl\"a, Finland
}
\author{M. Lebois}  
\affiliation{%
Institut de Physique Nucl\`eaire d'Orsay, 91406, Orsay, France
}
\author{L. Le Meur}  
\affiliation{%
 Subatech, IMT-Atlantique, Universit\'e de Nantes, CNRS-IN2P3, F-44307, Nantes, France
}
\author{T. Martinez}  
\affiliation{%
Centro de Investigaciones Energ\'eticas Medioambientales y Tecnol\'ogicas, E-28040, Madrid, Spain
}
\author{M. Monserrate}  
\author{A. Montaner-Piz\'a}  
\affiliation{%
 Instituto de F\'isica Corpuscular, CSIC-Universidad de Valencia, E-46071, Valencia, Spain
}
\author{I.D. Moore}  
\affiliation{%
 University of Jyv\"askyl\"a, FIN-40014, Jyv\"askyl\"a, Finland
}
\author{E. N\'acher}  
\affiliation{%
Instituto de Estructura de la Materia, CSIC, E-28006, Madrid, Spain
}
\author{S. E. A. Orrigo}  
\affiliation{%
 Instituto de F\'isica Corpuscular, CSIC-Universidad de Valencia, E-46071, Valencia, Spain
}
\author{H. Penttil\"a}  
\author{I. Pohjalainen}  
\affiliation{%
 University of Jyv\"askyl\"a, FIN-40014, Jyv\"askyl\"a, Finland
}
\author{A. Porta}  
\affiliation{%
 Subatech, IMT-Atlantique, Universit\'e de Nantes, CNRS-IN2P3, F-44307, Nantes, France
}
\author{J. Reinikainen}  
\author{M. Reponen}  
\author{S. Rinta-Antila}  
\affiliation{%
 University of Jyv\"askyl\"a, FIN-40014, Jyv\"askyl\"a, Finland
}
\author{B. Rubio}  
\affiliation{%
 Instituto de F\'isica Corpuscular, CSIC-Universidad de Valencia, E-46071, Valencia, Spain
}
\author{K. Rytk\"onen}  
\affiliation{%
 University of Jyv\"askyl\"a, FIN-40014, Jyv\"askyl\"a, Finland
}
\author{T. Shiba}  
\affiliation{%
 Subatech, IMT-Atlantique, Universit\'e de Nantes, CNRS-IN2P3, F-44307, Nantes, France
}
\author{V. Sonnenschein}
\altaffiliation[Present address: ]{%
 Faculty of Engineering, Nagoya University, Nagoya 464-8603, Japan
}
\affiliation{%
 University of Jyv\"askyl\"a, FIN-40014, Jyv\"askyl\"a, Finland
}
\author{A. A. Sonzogni}  
\affiliation{%
NNDC, Brookhaven National Laboratory, Upton, NY 11973-5000, USA
}
\author{E. Valencia}  
\affiliation{%
 Instituto de F\'isica Corpuscular, CSIC-Universidad de Valencia, E-46071, Valencia, Spain
}
\author{V. Vedia}  
\affiliation{%
Universidad Complutense, Grupo de F\'isica Nuclear, CEI Moncloa, E-28040, Madrid, Spain
}
\author{A. Voss} 
\affiliation{%
 University of Jyv\"askyl\"a, FIN-40014, Jyv\"askyl\"a, Finland
}
\author{J. N. Wilson}
\affiliation{%
Institut de Physique Nucl\`eaire d'Orsay, 91406, Orsay, France
}
\author{A. -A. Zakari-Issoufou} 
\affiliation{%
 Subatech, IMT-Atlantique, Universit\'e de Nantes, CNRS-IN2P3, F-44307, Nantes, France
}

\date{\today}

\begin{abstract}
In $\beta$-decay studies the determination of the decay probability to the ground state of the daughter nucleus often suffers from large systematic errors. The difficulty of the measurement is related to the absence of associated delayed $\gamma$-ray emission. In this work we revisit the $4\pi\gamma-\beta$ method proposed by Greenwood and collaborators in the 1990s, which has the potential to overcome some of the experimental difficulties. Our interest is driven by the need to determine accurately the $\beta$-intensity distributions of fission products that contribute significantly to the reactor decay heat and to the antineutrinos emitted by reactors. A number of such decays have large ground state branches. The method is relevant for nuclear structure studies as well. Pertinent formulae are revised and extended to the special case of $\beta$-delayed neutron emitters, and the robustness of the method is demonstrated with synthetic data. We apply it to a number of measured decays that serve as test cases and discuss the features of the method. Finally, we obtain
ground state feeding intensities with reduced uncertainty for four relevant decays that will allow future improvements in antineutrino spectrum and decay heat calculations using the summation method.
\end{abstract}

\maketitle

\section{Introduction
\label{intro}}

The measurement of ground state (g.s.)\ $\beta$-decay feeding probabilities is hampered by the absence
of associated $\gamma$ radiation. In $\beta^{-}$ decays the energy released is shared between
the electron and the antineutrino leading to continuous energy distributions,
extending from zero to the maximum decay energy $Q_{\beta}$.
This makes it difficult to determine precisely the number of $\beta$ particles.
The problem arises because of the difficulty of disentangling the featureless continuum associated with all of the decays to excited states from that to the g.s.\ through $\beta$-spectrum deconvolution.
In addition, electrons are easily absorbed or scattered by any material
surrounding the $\beta$ detector, and this effect must be properly taken into account 
in the response function of the $\beta$ detector. This explains why $\beta$-decay probabilities to the g.s.\ are often obtained indirectly.
   
The most common approach is to determine both the total number of decays and 
the number of decays proceeding to excited states, since the difference is due to 
decays to the g.s.. Usually the total number of decays is measured using a $\beta$ detector and 
the decays to excited states are obtained from high-resolution (HR) $\gamma$-ray spectroscopy 
and conversion electron spectroscopy 
to build the decay level scheme.  
Assigning the correct intensity for the decay to the g.s.\ is equivalent
to determining absolute $\gamma$ intensities.
The limited efficiency of HPGe detectors results in many weak transitions from
levels at high-excitation energy remaining undetected, the so-called \emph{Pandemonium} effect \cite{Har77}.
This shifts the apparent $\beta$ intensity, obtained from the 
intensity balance at each level, to levels at low-excitation energies. This in itself is not the real problem 
for g.s.\ feeding determination but the fact
that part of the missed  transitions can feed the g.s. directly, thus introducing a systematic error 
in the determination of absolute $\gamma$ intensities. 
In a strict sense the g.s.\ feeding probabilities obtained by this method should be considered as upper limits.


Greenwood and collaborators \cite{Gre92a} proposed the use of the \emph{Pandemonium}-free 
total absorption $\gamma$-ray spectroscopy (TAGS) technique~\cite{Rub05} in combination with
a $\beta$ detector to determine accurately the $\beta$ intensity to the g.s., 
in a way that will be explained later.
The method, termed the $4\pi\gamma-\beta$ method, was applied 
subsequently to determine the g.s.\ feeding probabilities for 34 fission 
products (FP) \cite{Gre95,Gre96}.

The TAGS technique aimed initially at the determination
of (relative) $\beta$ intensities to excited states.
It relies on the use of large close-to-$4\pi$ $\gamma$ calorimeters
made with scintillation material to detect the full de-excitation $\gamma$ cascade rather than
the individual transitions.
An ideal total absorption spectrometer would have 100\% $\gamma$-cascade detection efficiency 
and should be insensitive to $\beta$ particles.
It turns out that the TAGS technique can be used  to extract the g.s.\ $\beta$ intensity directly as is explained
below.

The first spectrometer designs 
emphasized the condition of insensitivity to $\beta$ particles, either by placing a $\beta$ detector outside 
the spectrometer \cite{Duk70} or placing a low-Z absorber material behind
the $\beta$ detector \cite{Byk80, Gre92b, Kar97} to minimize the penetration 
of the electrons or their bremsstrahlung radiation (in short $\beta$ penetration) 
in the scintillation volume. This had the undesirable effect of reducing
the $\gamma$-peak detection efficiency. However, the total detection efficiency
for $\gamma$ cascades of multiplicity 2 or higher remains close to 100\%
if the solid angle coverage is reasonably close to $4\pi$.
The rationale behind these initial designs is that $\beta$ penetration 
distorts the spectrometer $\gamma$ response by introducing a high energy tail.
The spectrometer response to decays is needed in the TAGS analysis of real spectrometers
to deconvolute the measured spectrum \cite{Tai07a}. 
The response must be obtained by Monte Carlo (MC)
simulations \cite{Can99} and the consensus at that time was 
that an accurate simulation of $\beta$-particle interactions
is more difficult than the simulation of $\gamma$ interactions.
Modern MC simulation codes like Geant4 \cite{Ago03}, have greatly improved the description
of low-energy electron interactions and provide a variety of tracking parameters to optimize
the simulation and improve the accuracy (see for example \cite{Wau09}). 
Thus the newest spectrometer designs
\cite{Tai11,DTAS_design,Kar16} do not make any special effort to minimize $\beta$ penetration 
and have a sizable response to g.s.\ decays. In this way the deconvolution of 
TAGS spectra provides, in a natural manner, the intensity of decays to the g.s. The first example of this was the decay of $^{102}$Tc for which the TAGS analysis~\cite{Alg10,Jor13}
confirmed the value of $92.9 (6)$\% quoted in ENSDF~\cite{Fre09} coming from HR spectroscopy. In the last years, many other examples have shown the potential of the TAGS technique to obtain g.s.\ feeding probabilities~\cite{Zak15,Val17,MTAS_neutrino_PRL,Fij17,100Tc,Simon_PRC, Gua19b, PRC_BDN}.
However, the TAGS deconvolution method has some limitations: 1) the difficulties in validating the shape of the MC simulations of the $\beta$ penetration in the spectrometer (see discussion in Ref.~\cite{Val17}) 2) the indeterminacy that can arise for particular decay intensity distributions, as will be shown later for the case of $^{103}$Tc, 3) the loss of sensitivity with decreasing g.s.\ $\beta$ intensity, 4) the difficulty of separating transitions to states at very low excitation energy, due to the limited energy resolution, and 5) the proper quantification of the systematic uncertainties.

In the present work we revisit the $4\pi\gamma-\beta$ method, which, as will be seen, is essentially free from problems 1) and 2) listed before and has different systematic uncertainties from the TAGS analysis. These differences mainly arise from the integral character of the $4\pi\gamma-\beta$ method (that uses the total number of counts in the spectra) that contrasts with the TAGS deconvolution, sensitive to the features of the shape of the experimental spectrum. As will be shown later, this minimizes the effect of the lack of knowledge on the precise de-excitation paths (a relevant source of systematic uncertainty in the TAGS analyses) in the results of the $4\pi\gamma-\beta$ method.  

Our interest in this topic arises from the importance of g.s.\ feeding probabilities for nuclear structure studies and reactor applications. In particular, it was recently renewed by the need to obtain accurate antineutrino energy spectra emitted by fission products (FP) using the $\beta$-intensity distributions $I_{\beta}  (E_{x})$ to weight the individual $\bar{\nu}_{e}$ spectra for each $\beta$ end-point $Q_{\beta}-E_{x}$. 
This is the basis of the summation method to obtain the spectrum of antineutrinos 
emitted by a nuclear reactor~\cite{summation_method}, which is calculated by weighting the spectrum
for each FP by the cumulative (or evolved individual) fission yield and the contribution of each fissile isotope. 
As it happens a number of fission products of significance in forming the reactor antineutrino spectrum have a strong or very strong g.s.\ decay branch \cite{Son15}. Some of them are $^{92}$Rb (95.2(7)$\%$), $^{96\text{gs}}$Y (95.5(5)$\%$), $^{142}$Cs (56(5)$\%$), $^{100\text{gs}}$Nb (50(7)$\%$), $^{140}$Cs (35.9(17)$\%$) or $^{93}$Rb (35(3)$\%$). Here the quoted g.s.\ feeding probability in brackets is coming from the Evaluated Nuclear Structure Data File (ENSDF)~\cite{ENSDF}. The importance of direct measurements of the g.s.\ feeding and the impact on antineutrino spectrum calculations can be illustrated with the example of $^{92}$Rb,
the top contributor above $E_{\bar{\nu_{e}}} = 5$~MeV~\cite{Zak15,Son15}, 
with $Q_{\beta}=8.095 (6)$~MeV \cite{AME16}.
The evaluated g.s feeding probability in the ENSDF database, based on HR spectroscopy, was $51(18)$\% 
\cite{Ban00} until 2012 when it changed to $95.2(7)$\% \cite{Ban12} as a result
of a new measurement of $\gamma$-ray intensities \cite{Lhe06}.  
From the deconvolution of the measured TAGS
spectrum we obtained a value of $87.5(25)$\% \cite{Zak15}, close to the last evaluation. 
This result was confirmed with the $91(3)$\% value obtained by the ORNL group (Fija{\l}kowska et al.) \cite{Fij17} using the MTAS total absorption spectrometer. In both cases, a significant improvement of reactor antineutrino summation calculations using the \textit{Pandemonium}-free value obtained with the TAGS technique was reported~\cite{Zak15, MTAS_neutrino_PRL}.

An accurate knowledge of the antineutrino spectrum is key to the analysis
of reactor antineutrino oscillation experiments. The standard method  
to obtain this spectrum is to apply a complex conversion procedure to integral $\beta$ spectra
for each of the main fissile isotopes in a reactor \cite{Sch81}. A re-evaluation of 
the conversion procedure \cite{Mue11,Hub11} led to the discovery of a deficit of about 6\% between
observed and estimated antineutrino fluxes \cite{Men11}. 
This was termed the reactor antineutrino anomaly.
Whether it indicates the existence of sterile neutrinos is 
a topic of very active investigation \cite{Vog15}.
The summation method allows an exploration of the origin of the anomaly
from a different perspective, and recently it was shown 
that the consistent inclusion of our newest TAGS decay data
reduces the discrepancy with the measured flux to the
level of the estimated uncertainties~\cite{Mag19}.
On the other hand the high statistics spectrum of detected antineutrinos 
obtained by the Daya Bay collaboration \cite{An16}
shows without doubt shape deviations with respect to the converted spectra. Such shape deviations were also seen by the Double Chooz~\cite{DoubleChooz} and Reno~\cite{RENO} collaborations.
The origin of this shape distortion is unclear but the use of summation
calculations allows one to explore a number of possibilities \cite{Hay15}.
Moreover, fine structure has been observed in the Daya Bay spectra
that has been ascribed to a few nuclear species with large g.s feeding on the
basis of summation calculations \cite{Son18}. This opens the
unlooked-for possibility of doing reactor $\bar{\nu}_{e}$ spectroscopy.

There is a related application in which the role of g.s.\ feeding  values is also of great relevance: the evaluation of the energy released in nuclear reactors by the radioactive decay of the FP, known as decay heat. The decay heat represents the dominant source of energy when a reactor is powered off and its proper determination is essential for safety reasons. It is usually evaluated by means of summation calculations that use the same ingredients mentioned above: fission yields, $\beta$-intensity distributions and $\beta$ spectra as a function of end-point energies to compute the evolution of the reactor decay heat with time. Some important decays for the determination of the reactor decay heat exhibit relevant g.s.\ $\beta$ branches (many of them are common cases with the reactor antineutrino spectrum explained before). The accurate determination of the decay heat is thus constrained by the availability of reliable g.s.\ feeding probability values.

This paper is organized as follows. The $4\pi\gamma-\beta$ method is presented
in Section \ref{method}, including a correction of the original formulae in Ref.~\cite{Gre92a}.
In addition a modification of the formulae for the case
of $\beta$-delayed neutron emitters is introduced. 
In the Appendix we provide a demonstration of the method using synthetic data
generated by realistic MC simulations.
The $4\pi\gamma-\beta$ method is applied to TAGS data taken at the 
Accelerator Laboratory of the University of Jyv\"{a}skyl\"{a} (JYFL-ACCLAB) in  
Section \ref{exper}. It gives a summary of relevant experimental details and presents
the results obtained, first for a number of relevant
test cases and then for a number of isotopes contributing significantly to reactor antineutrino spectra and decay heat. The g.s.\  intensities obtained
are compared with the TAGS deconvolution results and the literature. 
The last Section summarizes the conclusions.

\section{The $4\pi\gamma-\beta$ method
\label{method}}

The method is based on a comparison of the number of counts detected in
the $\beta$ detector $N_{\beta}$ and the number of counts registered in coincidence 
in both the $\beta$ detector and the total absorption spectrometer $N_{\beta \gamma}$. 
These can be written in terms of the number of decays $f_{i}$ feeding level $i$,
with $i=0$ representing the ground state, and they are related to the $\beta$ intensity
$I^{i}_{\beta}$ and the total number of decays $N_{d}$:

\begin{equation}
\begin{split}
N_{\beta} & = \varepsilon^{0}_{\beta} f_{0} + \sum_{i>0} \varepsilon^{i}_{\beta} f_{i} \\
&  = \varepsilon^{0}_{\beta} I^{0}_{\beta} N_{d} + \sum_{i>0} \varepsilon^{i}_{\beta} I^{i}_{\beta} N_{d}\\
N_{\beta \gamma} & = \varepsilon^{0}_{\beta \gamma} f_{0} + \sum_{i>0} \varepsilon^{i}_{\beta \gamma} f_{i} \\
& = \varepsilon^{0}_{\beta \gamma} I^{0}_{\beta} N_{d}+ \sum_{i>0} \varepsilon^{i}_{\beta \gamma} I^{i}_{\beta} N_{d}
\label{eq1}
\end{split}
\end{equation}

$\varepsilon^{i}_{\beta}$ is the probability
of detecting a signal in the $\beta$ detector for decays to level $i$ and $\varepsilon^{i}_{\beta \gamma}$
the probability of registering simultaneously signals in the $\beta$ detector and the total absorption $\gamma$-ray spectrometer. As seen in Eq.~\ref{eq1} we have separated explicitly the g.s.\ contribution. Let us define average $\beta$ efficiencies for decays to excited states only,
both in singles $\bar{\varepsilon}^{*}_{\beta}$ and in coincidence with the total absorption $\gamma$-ray spectrometer 
$\bar{\varepsilon}^{*}_{\beta \gamma}$

\begin{equation}
\begin{split}
\bar{\varepsilon}^{*}_{\beta} & 
=  \frac{\sum\limits_{i>0} \varepsilon^{i}_{\beta} f_{i}}{\sum\limits_{i>0} f_{i}} 
= \frac{\sum\limits_{i>0} \varepsilon^{i}_{\beta} I^{i}_{\beta}}{1-I^{0}_{\beta}} \\
\bar{\varepsilon}^{*}_{\beta \gamma} & 
= \frac{\sum\limits_{i>0} \varepsilon^{i}_{\beta \gamma} f_{i}}{\sum\limits_{i>0} f_{i}} 
=  \frac{\sum\limits_{i>0} \varepsilon^{i}_{\beta \gamma} I^{i}_{\beta}}{1-I^{0}_{\beta}} \\
\label{eq2}
\end{split}
\end{equation}

The reason for this somewhat artificial definition is that they are well determined from a TAGS analysis even in the specific cases when the spectrometer is insensitive to g.s.\ $\beta$ penetration
as will be shown later. Using these average efficiencies Eq.~\ref{eq1} can be rewritten as:

\begin{equation}
\begin{split}
\frac{N_{\beta}}{N_{d}} & 
=  \varepsilon^{0}_{\beta} I^{0}_{\beta} + \bar{\varepsilon}^{*}_{\beta} (1-I^{0}_{\beta}) \\
\frac{N_{\beta \gamma}}{N_{d}} & 
=  \varepsilon^{0}_{\beta \gamma} I^{0}_{\beta} + \bar{\varepsilon}^{*}_{\beta \gamma} (1-I^{0}_{\beta})
\label{eq3}
\end{split}
\end{equation}

\noindent from which $I^{0}_{\beta}$ can be determined:

\begin{equation}
 I^{0}_{\beta}
= \frac{\displaystyle 1 - \frac{ N_{\beta \gamma}}{ N_{\beta}} \frac{ \bar{\varepsilon}^{*}_{\beta}}{ \bar{\varepsilon}^{*}_{\beta \gamma}}}{\displaystyle 1+\frac{N_{\beta \gamma}}{ N_{\beta}} \frac{\varepsilon^{0}_{\beta}-\bar{\varepsilon}^{*}_{\beta}}{\bar{\varepsilon}^{*}_{\beta \gamma}} - \frac{\varepsilon^{0}_{\beta \gamma}}{\bar{\varepsilon}^{*}_{\beta \gamma}}} 
\label{eq4}
\end{equation}

This Eq.~\ref{eq4} can be compared with the equivalent one (Eq. 13) in the original
publication of Greenwood et al.\ \cite{Gre92a} that can be rewritten using our nomenclature
as:

\begin{equation}
 I^{0}_{\beta}
= \frac{\displaystyle 1 - \frac{ N^{*}_{\beta \gamma}}{ N_{\beta}} \frac{1}{ \bar{\varepsilon}^{*}_{\gamma}}}{\displaystyle 1+\frac{N^{*}_{\beta \gamma}}{ N_{\beta}} \frac{\varepsilon^{0}_{\beta}-\bar{\varepsilon}^{*}_{\beta}}{\bar{\varepsilon}^{*}_{\beta} \bar{\varepsilon}^{*}_{\gamma}}} 
\label{eq5}
\end{equation}

In the conversion we have used the following equivalence to the notation of Greenwood et al.:
$\bar{\varepsilon}^{*}_{\gamma} = 1-L$, $\varepsilon^{0}_{\beta} = f_{gs} \omega_{\beta}$,
and $\bar{\varepsilon}^{*}_{\beta} = f_{ex} \omega_{\beta}$.
Even assuming that the factorization 
$\bar{\varepsilon}^{*}_{\beta \gamma} =\bar{\varepsilon}^{*}_{\beta} \bar{\varepsilon}^{*}_{\gamma}$
is valid, there are differences between this expression and Eq.~\ref{eq4}.
A correction term is missing in the denominator and 
$N^{*}_{\beta \gamma}$ represents $N_{\beta \gamma}$ corrected by the $\beta$
penetration for g.s.\ decays (with probability $\varepsilon^{0}_{\beta \gamma}$)  and
for decays to excited states where the $\gamma$ cascade
is not detected in the spectrometer (with probability $\tilde{\varepsilon}^{i}_{\beta \gamma}$)
$N^{*}_{\beta \gamma} =  N_{\beta \gamma} - \varepsilon^{0}_{\beta \gamma} f_{0} 
- \sum_{i>0 } \tilde{\varepsilon}^{i}_{\beta \gamma} f_{i} $.
We show in the Appendix, using synthetic data, that Eq.~\ref{eq5} produces inconsistent results.

The application of the  $4\pi\gamma-\beta$ method 
requires the determination of the experimental ratio $R =  N_{\beta \gamma} / N_{\beta}$ 
and the estimation of three correction factors, $a$, $b$ and $c$, that are ratios of $\beta$ efficiencies

\begin{equation}
 I^{0}_{\beta}
= \frac{\displaystyle 1 - a R }{\displaystyle 1 + b R - c } 
\label{eq6}
\end{equation}

From its expression (compare Eq.~\ref{eq6} with Eq.~\ref{eq4})
one can see that  correction factor $a$ is close to (but larger than) one, 
correction factor $b$ is a small number and correction factor $c$ is a relative
measure of $\beta$ penetration for decays to the g.s.
To estimate accurately the correction factors we need to know the dependency of $\beta$
efficiency with endpoint energy and the $\beta$-intensity distribution with excitation energy (see Eq.~\ref{eq2}).
Notice that only the \emph{relative} $\beta$ intensity to excited states is required.
We also need to know the $\beta$-penetration probability in the total absorption $\gamma$-ray spectrometer.
Since $\gamma$ rays interact also in the $\beta$ detector we must take this effect into
account, which implies that we must have a knowledge of decay $\gamma$ cascades. These are also
needed to obtain the $\beta - \gamma$ detection efficiency. 
Conversion electrons are readily detected in the $\beta$ detector affecting both the $\beta$ counts
and the decay detection efficiency, and this is another effect that must be considered.
As a matter of fact all of this information
is required for the analysis of TAGS data  or is the result of such analysis (see Section \ref{exper}). The accuracy of the $4\pi\gamma-\beta$ method depends on the accuracy
of the ratio of counts $R$ and on the accuracy with which we
can determine the correction factors.
The integrated counts $N_{\beta}$ and $N_{\beta \gamma}$ can be obtained
from the measured $\beta$ and $\beta-\gamma$ spectra but corrections for contaminants
should be applied. The identification and quantification of  contaminants
is an important ingredient of the TAGS analysis, therefore providing 
the necessary information for the evaluation of this correction.
In summary the $4\pi\gamma-\beta$ method relies on the deconvolution of TAGS data
and becomes a natural extension of it.

The decay of $\beta$-delayed neutron emitters requires
special consideration. In this case the $\beta$-intensity distribution is the sum of two
contributions $I_{\beta} (E_{x}) = I_{\beta\gamma} (E_{x})+I_{\beta n} (E_{x})$. The first
one $I_{\beta\gamma} (E_{x})$ refers to decays that populate levels 
in the daughter nucleus that then de-excite
by emission of $\gamma$ rays. This is the one determined by the TAGS analysis.
The second one $I_{\beta n} (E_{x})$ refers to decays that populate 
levels above the neutron separation energy
$S_{n}$ which is then followed by the emission of one or more neutrons and eventually
$\gamma$ rays in the de-excitation of the final nucleus. 
This component can be obtained from the measured $\beta$-delayed neutron spectrum
and a knowledge of the branching probability to the different levels in the final
nucleus (see \cite{Val17} for further details). 
By separating the two components in the second row of Eq.~\ref{eq1} we obtain:

\begin{equation}
\begin{split}
N_{\beta \gamma} &  
= \varepsilon^{0}_{\beta \gamma} I^{0}_{\beta} N_{d}
+ \sum_{i>0} \varepsilon^{i}_{\beta \gamma} I^{i}_{\beta\gamma} N_{d}
+ \sum_{i>0} \varepsilon^{i}_{\beta n \gamma} I^{i}_{\beta n} N_{d}
\label{eq7}
\end{split}
\end{equation}

The last term represents the counts coming from the interaction of the 
$\beta$-delayed neutrons with the total absorption $\gamma$-ray spectrometer 
which is another source of contamination in the TAGS analysis
that must be corrected for as explained later.
After eliminating this contribution the second row in Eq.~\ref{eq1} becomes $N_{\beta\gamma}=\varepsilon^{0}_{\beta \gamma} I^{0}_{\beta} N_{d}+ \sum_{i>0} \varepsilon^{i}_{\beta \gamma} I^{i}_{\beta\gamma} N_{d}$. 
With a re-definition of the coincidence detection efficiency averaged over excited levels
(second row of Eq.~\ref{eq2}):

\begin{equation}
\begin{split}
\bar{\varepsilon}^{*}_{\beta \gamma} & 
=  \frac{\sum\limits_{i>0} \varepsilon^{i}_{\beta \gamma} I^{i}_{\beta\gamma}}{1-I^{0}_{\beta}} \\
\label{eq8}
\end{split}
\end{equation}

\noindent we arrive formally to the same formula Eq.~\ref{eq4} to calculate $I^{0}_{\beta}$.
Notice however that now $\bar{\varepsilon}^{*}_{\beta}$ is calculated with the
total $\beta$-intensity distribution $I^{i}_{\beta}$ while 
$\bar{\varepsilon}^{*}_{\beta\gamma}$ is calculated with the partial intensity distribution
$I^{i}_{\beta\gamma}$. The latter is normalized to $1-P_{n}$ instead of 1. The extension of the formulae for $\beta$-delayed neutron emitters was not discussed in the work of Greenwood \textit{et al.}.

The validation of the method with synthetic data is left for the Appendix. In the next section we apply the method to experimental data for a number of selected isotopes that either show some particularities in the use of the method or are important in determining the reactor antineutrino spectrum and/or the reactor decay heat.

\section{Experimental results
\label{exper}}

A campaign of TAGS measurements was carried out in 2014 at the upgraded Ion Guide Isotope Separator On-Line IGISOL IV facility~\cite{Moore_IGISOLIV} at the University of Jyv\"askyl\"a. One of the motivations for these measurements was to improve both reactor decay heat and antineutrino spectrum summation calculations, by providing data free from the \textit{Pandemonium} effect for some nuclei having significant g.s.\ feeding values. In the experiment we employed the 18-fold segmented NaI(Tl) Decay Total Absorption $\gamma$-ray Spectrometer (DTAS)~\cite{DTAS_design} in coincidence with a 3~mm-thick plastic scintillation detector. This $\beta$ detector was located at the centre of DTAS and in front of a movable tape for the implantation of the nuclei of interest and the removal of the daughter activity (see \cite{Gua16} for more details about the experiment). The mean efficiency of the $\beta$ detector is around 30$\%$ for endpoint energies above 2~MeV (see Fig.~\ref{fig1} in the Appendix for a similar detector), while the efficiency of DTAS for $\beta$-particles ranges from 8$\%$ at 3~MeV $\beta$ end-point energy to 44$\%$ at 8~MeV~\cite{Gua18}.

We provide in the following a brief description of TAGS data analysis for the reader's better understanding. The $\beta$-gated total energy deposited in DTAS was reconstructed off-line from the signals of the individual detector modules as described in~\cite{Gua18}, with threshold values of 90 keV for DTAS modules and 70 keV for the $\beta$ detector. The coincidence between DTAS and the $\beta$ detector allowed us to get rid of the environmental background. Other sources of contamination need to be accounted for. These include in general the activity of the descendants. For each descendant that
contributes significantly we measure the shape of its energy spectra or, in the case of well known decays, we obtain it through MC simulations using the available decay data. If possible the
normalization of these spectra is obtained by adjustment to salient 
features on the measured parent decay spectra that can be identified as due to
the descendant activity, otherwise from the relation of parent-descendant half-lives.
In the case of $\beta$-delayed neutron emitters, as was mentioned above,
the $\beta$-delayed neutron branch introduces an additional contamination
that includes the interaction of neutrons with DTAS. The shape of the contaminant
spectrum is obtained by MC simulation following a special
procedure detailed in \cite{Val17} and the normalization is
obtained by adjustment to the measured spectra, if possible, otherwise it is given
by the $P_{n}$ value. We also take into account the electronic pulse summing-pileup 
effect that contributes to the  distortion of the spectra. The need to
consider two components (summing and pileup) is particular to multi-detector systems.
The pileup originates in the superposition of different event signals in the same detector module 
within the analog-to-digital converter (ADC) time gate. 
The summing is due to the sum of signals corresponding to different events 
that are detected in different modules within the same ADC gate. 
The summing-pileup contribution is calculated  
by means of a MC sampling method~\cite{Gua18}
specifically developed for segmented spectrometers. 
The calculated spectrum is normalized from the detection rate
and the length of the ADC gate~\cite{Gua18}.

In order to determine the $\beta$ intensities from TAGS experimental spectra, we followed the method developed by the Valencia group~\cite{Can99,Tai07a,Tai07b}. For this, one has to solve the inverse problem $d_{i}= \sum_{j} R_{ij}(B) f_{j} + C_{i}$, where $d_{i}$ represents the number of counts in channel $i$ of the spectrum, $f_{j}$ is the number of events that feed level $j$ in the daughter nucleus, $R_{ij}$ is the response function of the spectrometer, which depends on the branching ratios ($B$) for the different de-excitation paths of the states populated in the decay, and $C_{i}$ is the sum of all contaminants at channel $i$. 

To build the spectrometer response to decays we need the response to
individual $\gamma$ rays and $\beta$ particles \cite{Can99}. 
These are obtained from MC simulations using a very detailed description 
of the measurement setup (including electronic thresholds), 
carefully benchmarked with laboratory sources. 
We also need the branching ratio matrix describing the de-excitation
pattern as a function of level excitation energy, including the conversion electron process. 
This is obtained from the HR spectroscopy level scheme at low excitation energies
supplemented with the predictions of the  Hauser-Feshbach nuclear statistical 
model above a given excitation energy where the levels are treated 
as a binned continuum \cite{Tai07b}.
The statistical model provides a realistic description of the electromagnetic cascade
energy and multiplicity distribution, that in modern segmented spectrometers
can be tested as well \cite{Gua19a,Gua19b,PRC_BDN} and eventually modified.

Finally the TAGS spectrum deconvolution is carried out by applying a suitable algorithm, which in the present case is the expectation maximization (EM) algorithm, to extract the $\beta$-feeding distribution \cite{Tai07a}.

In order to apply the $4\pi\gamma-\beta$ method to obtain $I^0_{\beta}$ 
we must determine the experimental number of counts $N_{\beta\gamma}$ and $N_{\beta}$.
$N_{\beta\gamma}$ is obtained from the number of counts in the $\beta$-gated 
DTAS spectrum after correction for the counts due to the contaminants. 
As we mentioned in Section~\ref{method} this correction follows closely the one 
applied to TAGS spectra for deconvolution since the contamination counts are determined 
by integration of  the corresponding TAGS contamination spectra that we just described. 
In this line, the counts due to the activity of the descendants and, if needed, the $\beta$-delayed neutron
branch contribution are subtracted.
In the case of the summing-pileup contribution the counts are $added$ 
since each count in the summing-pileup spectra represents the loss of two
events. The uncertainty of $N_{\beta\gamma}$ is estimated by considering 
the uncertainties in the normalization factors of the different components,
taken from the TAGS analysis. Note that in all cases we integrate the full $\beta$-gated DTAS spectra, since experimental thresholds are already taken into account when requiring the $\beta$-gating condition, as mentioned above. The experimental thresholds are also taken into account in the MC efficiencies employed for the determination of coefficients $a$, $b$ and $c$ of Eq.~\ref{eq6}. 

$N_{\beta}$ is calculated as the number of counts in the spectrum of the $\beta$ plastic detector above the threshold without any coincidence condition. 
In addition to the counts due to descendants, which are estimated from the TAGS analysis, in this case we also need to subtract environmental background counts, although this is a small amount. They are obtained from measurements without beam and normalized by the relative
measurement times. The counts lost by electronic pulse pileup are added to the result.
In this case, since we are dealing with a single detector,
the calculation and normalization of the pileup contribution is performed as in Ref.~\cite{TAS_pileup}.
The uncertainty on $N_{\beta}$ comes from the uncertainties in the normalization factors 
of the contaminants.
For the environmental background we take a 20$\%$ uncertainty, which is the maximum deviation observed in tests with laboratory sources,
while for the rest of the contaminants we take the same uncertainties used for the TAGS analysis. 

Finally we should mention that in general the ratio $N_{\beta\gamma}/N_{\beta}$  needs to be corrected 
for differences in the data acquisition dead-times.
In the present case this is not necessary because of the way our acquisition system works: every acquisition
channel is gated with a common gate signal which is an OR of all individual detector triggers.

The results of the application of the $4\pi\gamma-\beta$ method 
to part of the data obtained in the 2014 measurement campaign are presented below. Firstly cases of particular interest showing how the method works (sections~\ref{exper103} and~\ref{experOther}) and secondly (section~\ref{experNeutrino}) cases that are of relevance for reactor antineutrino summation calculations. They are summarized in Table~\ref{Table_GS_values}.

\begin{table}
\begin{ruledtabular}
\begin{tabular}{cccc}
 Isotope & \multicolumn{3}{c}{$I^0_{\beta}$ [$\%$]}  \\
 \cmidrule(r){2-4}
 &  ENSDF & TAGS & $4\pi\gamma-\beta$  \\  \hline
 \rule{0pt}{3ex}
$^{95}$Rb & $\leq$0.1 & $0.03_{-0.02}^{+0.11}$ & -0.2(42)\\
\rule{0pt}{3ex}
$^{100\text{gs}}$Nb  & 50(7) & $46_{-15}^{+16}$ & 40(6)\\
\rule{0pt}{3ex}
$^{102\text{m}}$Nb  & - & $42.5_{-10.0}^{+9.3}$ & 44.3(28)\\ 
\rule{0pt}{3ex}
$^{100}$Tc  & 93.3(1) & 93.9(5) & 92.8(5)\\
\rule{0pt}{3ex}
$^{103}$Tc & 34(8) & - & $45.6_{-0.9}^{+1.5}$\\
\rule{0pt}{3ex}
$^{137}$I  & 45.2(5) & $50.8_{-4.3}^{+2.7}$ & 45.8(13)\\
\rule{0pt}{3ex}
$^{140}$Cs  & 35.9(17) & 39.0$_{-6.3}^{+2.4}$ & 36.0(15)\\ 
\end{tabular}
\caption{\label{Table_GS_values}Ground state feeding intensities for a few decays of interest. The values obtained by the $4\pi\gamma-\beta$ method are compared with results from the TAGS analysis and from evaluations in the ENSDF database~\cite{ENSDF}.}
\end{ruledtabular}
\end{table}

\subsection{The case of $^{103}$Tc
\label{exper103}}

\begin{figure}[h]
 \begin{center}
 \includegraphics[width=8.6cm]{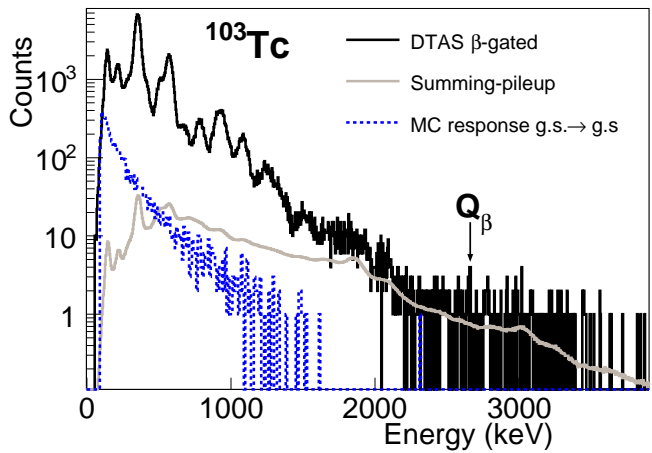}
 \caption{(Color online) Experimental $\beta$-gated spectrum (solid black) compared with the MC response of DTAS for the transition to the g.s.\ of $^{103}$Ru (dotted blue). The normalized summing pileup contribution is also shown (grey).}
 \label{fig2}
 \end{center}
\end{figure}

\begin{figure}[h]
 \begin{center}
 \includegraphics[width=8.6cm]{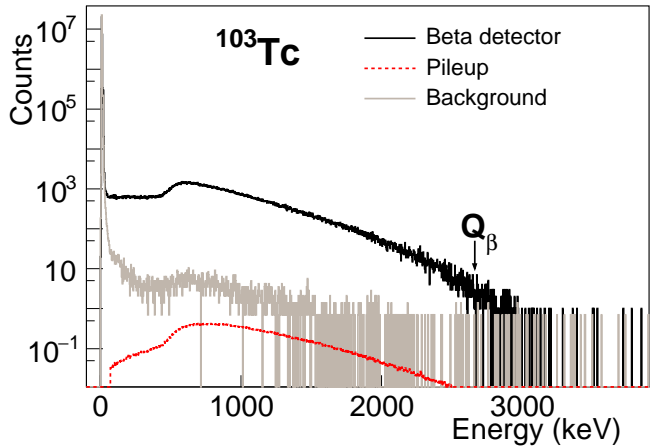} 
 \caption{Spectrum in singles of the plastic $\beta$ detector for the decay of $^{103}$Tc (solid black line). The contaminants are shown with the appropriate normalization.}
 \label{fig3}
 \end{center}
\end{figure}

One of the cases studied is the decay of the $5/2^+$ g.s.\ of $^{103}$Tc 
into $^{103}$Ru with a $Q_{\beta}$ value of 2663(10)~keV~\cite{AME16}. This TAGS measurement was assigned first priority for the prediction of the reactor decay heat with U/Pu fuel and second priority for Th/U fuel by the IAEA~\cite{IAEA}. In addition to the $3/2^+$ g.s.\ of the daughter nucleus, the decay also populates a $5/2^+$ state at only 2.81(5)~keV excitation energy, 
as observed in a previous HR spectroscopy experiment~\cite{Niizeki103Tc}. 
Since we are not able to separate the two close-lying states we refer to their summed decay intensity as the g.s.\ intensity $I^{0}_{\beta}$.
The DTAS and $\beta$-detector spectra for this measurement are shown 
in Fig.~\ref{fig2} and Fig.~\ref{fig3} respectively, together with the spectra of the
contaminants. Since  $^{103}$Ru is very long lived ($T_{1/2}=39.2$~days) these are limited to
the summing-pileup in the first case and the pileup and environmental background
in the second case.

\begin{figure}[h]
 \begin{center}
 \includegraphics[width=8.6cm]{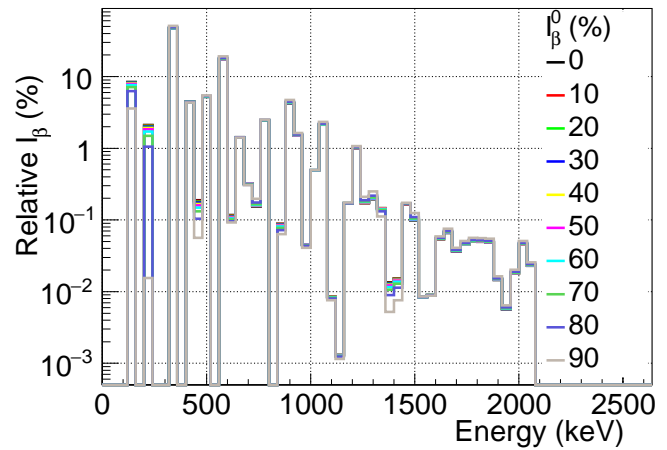}
 \caption{(Color online) Relative $\beta$ intensities populating the excited states of $^{103}$Ru in the decay of $^{103}$Tc. The $\beta$ intensities obtained from the TAGS analysis for different fixed values of $I^0_{\beta}$ are normalized to 1-$I^0_{\beta}$.}
 \label{fig4}
 \end{center}
\end{figure}

\begin{figure}[h]
 \begin{center}
 \includegraphics[width=8.6cm]{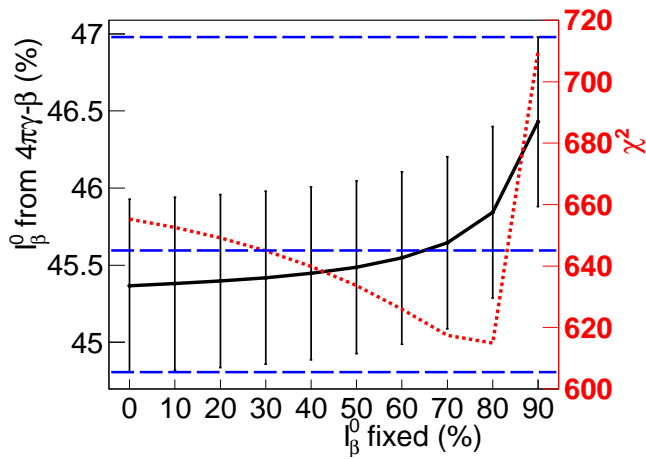}
 \caption{(Color online) $I^0_{\beta}$ values obtained with the $4\pi\gamma-\beta$ method for different TAGS analyses performed with $I^0_{\beta}$ fixed to values between 0 and 90 $\%$ (solid black). The $\chi^2$ of each TAGS analysis is shown in dotted red. The horizontal dashed blue line shows the mean $I^0_{\beta}$ value. The upper and lower limits considered for the uncertainty are represented by horizontal dotted gray lines.}
 \label{fig5}
 \end{center}
\end{figure}

It turns out that $^{103}$Tc is a special case. In the TAGS analysis presented in~\cite{Acta103}, we found that the reproduction of the measured DTAS spectrum is almost insensitive to the value of the g.s.\ feeding intensity, which is an unusual situation. To illustrate this in Fig.~\ref{fig4} we show the relative $\beta$ intensities to excited states of $^{103}$Ru when the g.s.\ $\beta$ intensity is fixed in the TAGS analysis to values between 0\% and 90\% in steps of 10\%. As can be observed they are almost unchanged in the range 0\% to 80\%. For a value of 90$\%$ there is a sizable effect on the $\beta$ intensity to states below 300~keV. The different $I^0_{\beta}$ values introduce a change of around 15$\%$ in the $\chi^2$ of the TAGS analysis (see Fig.~\ref{fig5}), with a minimum at $70-80$\%. In other words, in this particular case one should not trust the TAGS analysis to obtain the g.s.\ feeding probability.
The reason for this insensitivity is to be found in the shape of the response for the g.s.\ to g.s.\ transition. As shown in Fig.~\ref{fig2} the energy dependence of the MC simulated response for this transition happens to be similar to the overall shape of the total absorption experimental spectrum, thus the deconvolution algorithm is not very sensitive to the g.s.\ contribution.

We turn now to the $4\pi\gamma-\beta$ method. We obtain a ratio of counts $R$
for the decay of $^{103}$Tc  of 0.495(5). We use Eq.~\ref{eq4} to obtain $I^0_{\beta}$ using correction factors calculated with the $\beta$-intensity distributions shown in Fig.~\ref{fig4}. As in the case of the synthetic data (see Appendix), we obtain very stable results with the $4\pi\gamma-\beta$ method. In Fig.~\ref{fig5} the $I^0_{\beta}$ values obtained with this method for each fixed $I^0_{\beta}$ in the TAGS analysis are presented, showing very small variations when the values vary between 0\% and 90\%. The uncertainties for each $I^0_{\beta}$ determined with the $4\pi\gamma-\beta$ method are obtained from the uncertainty in the ratio of the number of counts $R$, which combines statistical uncertainties and uncertainties in the correction for contaminations, and the uncertainties in the correction factors. We use a set of $\beta$-intensity distributions, a total of 17 solutions of the TAGS analysis compatible with the experimental data, to obtain different correction factors and evaluate the dispersion of $I^0_{\beta}$ values determined with the $4\pi\gamma-\beta$ method. It was found to be very small, which is related to the fact that the uncertainties in the average efficiencies defined in Eq.~\ref{eq2} are small, as discussed in the Appendix. As explained in detail in \cite{Val17} each of these $\beta$-intensity distributions obtained in the TAGS
analysis takes into account the effect of several systematic uncertainties related to the branching ratio matrix
used to build the spectrometer response, to the accuracy of MC simulations, 
to the normalization of contaminants and even to the deconvolution method. We would like to note that errors quoted in Table~\ref{Table_GS_values} for the $4\pi\gamma-\beta$ values are the quadratic sum of the uncertainty in the calculation of the ratio of counts $R$ (the dominant one) and the small contribution coming from the application of the method with all solutions used to estimate the error budget of the TAGS analyses. 

The mean of all results in Fig.~\ref{fig5} gives a value of $45.6_{-0.9}^{+1.5}$\%. The uncertainty is a conservative estimate based on the lowest and highest uncertainty deviations. This result is compatible with the value 41(10)$\%$ obtained in the HR spectroscopy work of Niizeki \textit{et al.}~\cite{Niizeki103Tc}. However, it is larger than the 34(8)$\%$ value reported in the ENSDF evaluation~\cite{NDS_103} based also on HR spectroscopy studies. The difference between both is related to the adopted intensity of the 346.4~keV $\gamma$ ray used for normalization: Niizeki \textit{et al.} uses an intensity of 16$\%$ for this $\gamma$ ray, whereas the ENSDF evaluation uses a value of 18.4$\%$ obtained in a fission yield measurement~\cite{fissionYield_103Tc}. 
This example shows one of the difficulties faced when assigning g.s.\ feeding probabilities in HR spectroscopy.

\begin{figure}[h]
 \begin{center}
 \includegraphics[width=8.6cm]{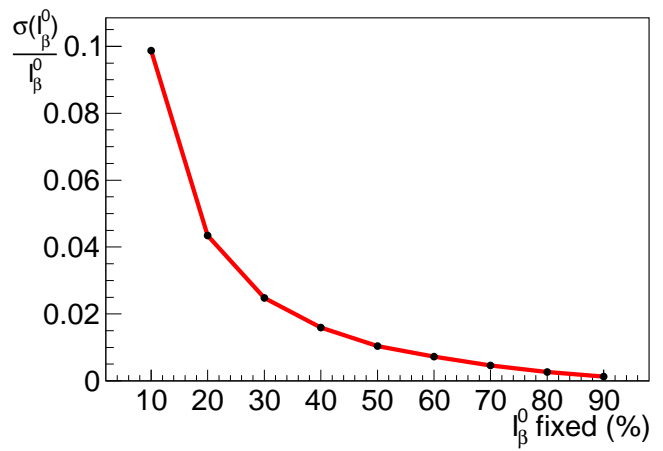}
 \caption{Relative uncertainty of the $I^0_{\beta}$ values obtained with the $4\pi\gamma-\beta$ method applied to MC simulations.  The TAGS results obtained with $I^0_{\beta}$ fixed to different values from 10 to 90 $\%$ in the analysis of the decay of $^{103}$Tc have been used as input for the event generator employed in the simulations. The uncertainty in the ratio of counts $R$ was kept fixed at the experimental 1$\%$ value.}
 \label{fig6}
 \end{center}
\end{figure}

We also use the decay of $^{103}$Tc to perform an illustrative exercise showing the dependence of the uncertainty of $I^0_{\beta}$ in the $4\pi\gamma-\beta$ method with $I^0_{\beta}$ for a fixed value of the uncertainty in $R$. The $\beta$-intensity distributions obtained in the TAGS analysis with fixed $I^0_{\beta}$ values between 10\% and 90\% (shown in Fig.~\ref{fig4}) have been used as input to MC simulations of DTAS and plastic scintillation spectra (see also the Appendix). The $4\pi\gamma-\beta$ method was then applied to these spectra. The uncertainty in the ratio of counts $R$ was fixed to 1\%, the value of the current measurement. As shown in Fig.~\ref{fig6}, the relative error in the determination of the g.s.\ feeding probability with the $4\pi\gamma-\beta$ method varies between 10\% at $I^0_{\beta}=10\%$ and 1\% at $I^0_{\beta}=90\%$. Thus the precision of the method is severely limited by statistics at low values of the g.s.\ feeding intensity.

\subsection{Cases with extreme values of the g.s.\ feeding intensity
\label{experOther}}

The $4\pi\gamma-\beta$ method has been applied to other test cases measured in the same DTAS experimental campaign. In those cases the TAGS analyses did show a sensitivity to the g.s.\ to g.s.\ transition, thus allowing us to compare the $I^0_{\beta}$ determined from the deconvolution with the value obtained by means of the $4\pi\gamma-\beta$ counting method presented here.
In particular two cases are included here due to the extreme character of their g.s.\ feeding probability and importance: the decay of $^{95}$Rb, a $\beta$-delayed neutron emitter where the first forbidden $5/2^{-}\rightarrow1/2^+$ g.s.\ to g.s.\ transition is hindered, and the decay of $^{100}$Tc, dominated by the large Gamow-Teller $1^{+}\rightarrow0^+$  g.s.\ to g.s.\ branch. 

In the first case, the decay of $^{95}$Rb, we obtain an almost zero g.s.\ to g.s.\ feeding from TAGS analysis, in agreement with previous HR spectroscopy measurements~\cite{PRC_BDN} (see Table~\ref{Table_GS_values}). Nevertheless, our TAGS analysis shows that the HR data is affected by a strong \emph{Pandemonium} effect~\cite{PRC_BDN}. The $4\pi\gamma-\beta$ method determines also a $I^0_{\beta}$ value that is almost zero, though the relative uncertainty is large, as can be expected from the discussion in the previous Section. In fact (see Table~\ref{Table_GS_values}) in this case the uncertainty is much larger than that determined by TAGS spectrum deconvolution. Due to the relatively small fission yield of $^{95}$Rb, the impact of our TAGS results in reactor antineutrino spectrum calculation is less than $1\%$ between 7 and 9~MeV. For the same reason, in spite of being assigned first priority for the U/Pu fuel decay heat~\cite{IAEA}, the impact of these TAGS results is also subpercent on the electromagnetic component of the reactor decay heat of $^{235}$U and $^{239}$Pu for times shorter than 1~s~\cite{PRC_BDN}.

In the second case, the decay of $^{100}$Tc, of interest for nuclear structure, a large $I^0_{\beta}$ value of 93.9(5)$\%$ is determined from the TAGS spectrum deconvolution. As described in \cite{NIMA_Vase} a different $\beta$ detector was employed in this measurement, which consists of a vase-shaped thin plastic scintillator with  close-to-$4\pi$  solid angle coverage. The value of $I^0_{\beta}$ obtained with the TAGS technique is compatible with the previous value from HR measurements (see Ref.~\cite{100Tc} for a detailed discussion).  The $4\pi\gamma-\beta$ method gives a value of 92.8(5)$\%$ in agreement with the value from the TAGS analysis, thus confirming this important result. In this case the value quoted by ENSDF is in agreement within the uncertainties with both results. The $\beta$-intensity distribution of $^{100}$Tc decay serves as a benchmark for theoretical estimates of the nuclear matrix elements (NME) in the A=100 system that enter into the calculation of the double $\beta$-decay process in $^{100}$Mo. NME represent the largest uncertainty in the half-life estimate of the neutrino-less branch, thus limiting our ability to extract information on this process beyond the Standard Model.

\subsection{Reactor antineutrino spectrum cases
\label{experNeutrino}}

The remaining cases presented here are decays of fission fragments contributing significantly to the reactor antineutrino spectrum: $^{100\text{gs}}$Nb, $^{102\text{m}}$Nb, $^{137}$I and $^{140}$Cs. Table~\ref{Table_nubar_cases} provides the percent contribution of the four isotopes to the total antineutrino spectrum for both $^{235}$U and $^{239}$Pu fission in three $E_{\bar{\nu}_{e}}$ energy intervals covering the range from 3 to 6~MeV. These percentages
were calculated using the Nantes summation 
method~\cite{Zak15}. All cases listed in Table~\ref{Table_nubar_cases} have been assigned a first priority for TAGS measurement in the IAEA report~\cite{IAEA}, while $^{137}$I and $^{100\text{gs}}$Nb are also considered high-priority cases for the reactor decay heat by the IAEA~\cite{IAEA}. 

\begin{table}
\begin{ruledtabular}
\begin{tabular}{ccccccc}
Isotope & \multicolumn{2}{c}{3-4~MeV [\%]} & \multicolumn{2}{c}{4-5~MeV [\%]} &
\multicolumn{2}{c}{5-6~MeV [\%]} \\
\cmidrule(r){2-3} \cmidrule(r){4-5} \cmidrule(r){6-7}
& $^{235}$U & $^{239}$Pu & $^{235}$U & $^{239}$Pu & $^{235}$U & $^{239}$Pu \\  \hline
\rule{0pt}{3ex}
\rule{0pt}{3ex}
$^{100\text{gs}}$Nb & 3.5 & 4.5 & 5.3 & 7.6 & 5.8 & 9.0\\
\rule{0pt}{3ex}
$^{102\text{m}}$Nb & 0.7 & 1.5 & 0.7 & 1.7 & 0.4 & 1.0 \\
\rule{0pt}{3ex}
$^{137}$I & 1.7 & 1.6 & 2.2 & 2.3 & 2.0 & 2.3 \\
\rule{0pt}{3ex}
$^{140}$Cs & 2.8 & 2.9 & 3.3 & 3.7 & 2.5 & 3.0 \\
\end{tabular}
\caption{\label{Table_nubar_cases} Contribution in $\%$ of the selected cases to the reactor
antineutrino spectra of $^{235}$U and $^{239}$Pu at different energy ranges (based on the
Nantes summation method~\cite{Zak15}).}
\end{ruledtabular}
\end{table}

As can be observed in Table~\ref{Table_GS_values} the relative uncertainty in $I^0_{\beta}$ obtained by TAGS spectrum deconvolution is rather large in these four cases. In particular in the case of $^{100\text{gs}}$Nb, estimated to be one of the largest contributors in the region of the spectral distortion around 5~MeV, reaches 35\%. In the case of $^{102\text{m}}$Nb it is 24\%. In both cases the TAGS analysis is strongly affected by the uncertainty in the contamination of the parent activity (see Ref.~\cite{Gua19b} for more details). The characteristic of the $4\pi\gamma-\beta$ method of being almost insensitive to the actual $\beta$-intensity distribution obtained in the TAGS analysis can be of advantage here.

As can be seen in Table~\ref{Table_GS_values} an overall good agreement is found between the g.s.\ feeding probabilities obtained with the $4\pi\gamma-\beta$ method and those determined in the TAGS analyses. The $4\pi\gamma-\beta$ method, however, produces results 
with much smaller relative uncertainties 
compared to the TAGS analysis for the two Nb cases: 15\% and 6\% respectively. Smaller uncertainties
are also obtained for $^{137}$I and $^{140}$Cs.
The central values are in agreement within uncertainties for both methods. However,
observing all the values in the Table~\ref{Table_GS_values} one could also claim that the $4\pi\gamma-\beta$ method tends to produce results systematically smaller than TAGS spectrum deconvolution, with the exception of $^{102\text{m}}$Nb. Whether this is true and could be related to some systematic error in one of the two methods should be studied further.

Compared to the values in the ENSDF database~\cite{ENSDF} we observe (see Table~\ref{Table_GS_values}) that the 
$4\pi\gamma-\beta$ method is in close agreement for $^{137}$I and $^{140}$Cs, and 20\% smaller for 
$^{100\text{gs}}$Nb although in agreement within  uncertainties. No value is available for 
$^{102\text{m}}$Nb in the ENSDF database.

\section{Summary and conclusions
\label{summary}}

In this work we have addressed the determination of the $\beta$-decay intensity to the g.s.\ of the daughter nucleus $I^0_{\beta}$ by means of a $\beta$-$\gamma$ counting method. This approach, initially proposed by Greenwood \textit{et al.}~\cite{Gre92a}, relies on the use of a high-efficiency $\gamma$ calorimeter in coincidence with a $\beta$ detector. The original $4\pi\gamma-\beta$ method has been revised and some inconsistencies in the formulae were found and corrected. Furthermore we extended the formulae to the particular case of $\beta$-delayed neutron emitters, to take into account the fraction of decays proceeding by neutron emission. We have shown that the method becomes an extension of and relies on the total absorption $\gamma$-ray spectroscopy technique. The analysis performed using this technique provides the information needed to calculate the quantities required by the $4\pi\gamma-\beta$ method as well as the accurate determination of contaminant contributions, which results in an improved overall accuracy. The robustness of the method is demonstrated
using synthetic decay data obtained from MC simulations. It was shown that statistics becomes a limiting factor for determining with precision the decay probability to the g.s.\  as this probability becomes smaller. 

We have applied the $4\pi\gamma-\beta$ method to a number of cases measured in our last experimental campaign with the DTAS spectrometer at the IGISOL IV facility. The main goal of the campaign was to measure accurately the $\beta$-intensity distribution in the decay of FP of importance in determining the antineutrino spectrum and the decay heat from reactors, several of which have a large decay to the g.s. The TAGS analysis of one case, $^{103}$Tc, turned out to be insensitive to the value of the g.s.\ feeding probability, and the  $4\pi\gamma-\beta$ counting method was the only way to determine its rather large value of about 45\%. Even though $^{103}$Tc is a special case, this shows one of the potential issues when determining the g.s.\ $\beta$-decay intensity from the deconvolution method. For the remaining cases, with $I^0_{\beta}$ values ranging from 0 to more than 90$\%$, good agreement between the g.s.\ feeding probabilities determined in the TAGS analysis and those obtained with the $4\pi\gamma-\beta$ method was observed. This provides a confirmation of previous g.s.\ feeding probabilities obtained by the TAGS deconvolution method and in particular confirms the accuracy of the simulation of the shape of the $\beta$ penetration, to which the $4\pi\gamma-\beta$ method is not sensitive. For the cases studied we found that, with the exception of the negligible intensity of $^{95}$Rb, the uncertainties in the $4\pi\gamma-\beta$ method are smaller. Besides case-specific reasons this is related to the small effect of our lack of knowledge of level de-excitations in the daughter nucleus on this method, whilst it represents a significant fraction of the error budget in the TAGS analysis. 
In particular the uncertainties for the important contributors to the reactor antineutrino spectrum $^{100\text{gs}}$Nb and $^{102\text{m}}$Nb are reduced by factors 2.5 and 4, respectively, resulting in more precise antineutrino spectra for these nuclei with a corresponding improvement in future summation calculations.

In conclusion, the $4\pi\gamma-\beta$ method represents an alternative, generally superior, approach to the TAGS spectrum deconvolution to determine g.s.\ feeding probabilities. The potential of this tool to provide accurate and precise $I^0_{\beta}$ values, which is hampered by the lack of associated $\gamma$-ray emission, was demonstrated in this work. Ground state feeding probabilities are needed to determine the absolute value of the decay intensity to excited states and carry important information on the nuclear structure. In addition, due to the significant influence of the $\beta$-decay branches to the g.s.\ on
the reactor antineutrino spectrum and decay heat, our capacity to better determine such transitions will help us understand the challenging puzzle of reactor antineutrinos, while improving decay heat predictions.

\begin{acknowledgments}

This work has been supported by the Spanish Ministerio de Econom\'ia y Competitividad under Grants No. FPA2011-24553, No. AIC-A-2011-0696, No. FPA2014-52823-C2-1-P, No. FPA2015-65035-P, No. FPI/BES-2014-068222, No. FPA2017-83946-C2-1-P, No. RTI2018-098868-B-I00 and the program Severo Ochoa (SEV-2014-0398), by the Spanish Ministerio de Educaci\'on under the FPU12/01527 Grant, by the European Commission under the FP7/EURATOM contract 605203, the FP7/ENSAR contract 262010, the SANDA project funded under H2020-EURATOM-1.1 contract 847552, the Horizon 2020 research and innovation programme under grant agreement No. 771036 (ERC CoG MAIDEN), by the Generalitat Valenciana regional funds PROMETEO/2019/007/, and by the $Junta~para~la~Ampliaci\acute{o}n~de~Estudios$ Programme (CSIC JAE-Doc contract) co-financed by ESF. We acknowledge the support of the UK Science and Technology Facilities Council (STFC) Grant No. ST/P005314/1 and of the Polish National Agency for Academic Exchange (NAWA) under Grant No. PPN/ULM/2019/1/00220. This work was also supported by the Academy of Finland under the Finnish Centre of Excellence Programme (Project No. 213503, Nuclear and Accelerator-Based Physics Research at JYFL). A.K. and T.E. acknowledge support from the Academy of Finland under Projects No. 275389 and No. 295207, respectively. This work has also been supported by the CNRS challenge NEEDS
and the associated NACRE project, the CNRS/in2p3 PICS TAGS between Subatech and IFIC, and the
CNRS/in2p3 Master projects Jyvaskyla and OPALE.

\end{acknowledgments}

\appendix

\section*{Appendix
\label{app}}

\subsection*{Application to synthetic data}
The synthetic data emulate the decay of  $^{87}$Br, $^{88}$Br, and $^{94}$Rb, 
that we have previously investigated with the TAGS technique \cite{Val17}.
From the deconvolution of TAGS spectra we obtained ground state feeding
intensities of $10.10^{+1.19}_{-0.94}$\% and $4.72^{+1.03}_{-2.19}$\% 
for $^{87}$Br and $^{88}$Br respectively.
By comparison, ENSDF assigns a decay intensity to the g.s.\ of $12.0(19)$\%
in the first case \cite{Joh15} and an upper limit of 11\% in the second case \cite{Mcc14}.
The $\mathrm{g.s.} \rightarrow \mathrm{g.s.}$  decay in $^{94}$Rb corresponds 
to a third forbidden transition with negligible intensity. 
These three nuclei are $\beta$-delayed neutron emitters with
neutron emission probabilities of $2.43(14)$\%, $6.75(18)$\% and $10.24(21)$\%, respectively~\cite{IAEAbDN}.
For simplicity we have not included this decay channel in the
simulation since we are interested here in testing the performance of the method. 
Accordingly only one $\beta$-intensity distribution (normalized to 1)
is used to calculate the correction factors in Eq.~\ref{eq4} (Section~\ref{method}) and
we need to scale by $1-P_{n}$ the $I^{0}_{\beta}$ obtained 
in order to compare with the true value, defined as the input value for the simulation.

The measurement was performed using a compact 12-fold segmented 
$\mathrm{BaF}_{2}$ total absorption spectrometer with cylindrical shape
and a thin Si detector placed close to the source
position subtending a solid angle fraction of $\approx30$\% as described in \cite{Val17}.
Spectra were simulated with the Geant4 Simulation Toolkit \cite{Ago03} 
implementing the detailed description of the setup and
using decay cascades 
produced by the DECAYGEN event generator \cite{Tai07b}.
The inputs to the event generator are the branching ratio matrix used in the 
TAGS analysis and the resulting $\beta$-intensity distribution that was adopted
as the final solution in \cite{Val17}. 
This provides a very realistic simulation of the decay and its detection.
In the simulation, as well as in the analysis,
we assume that all $\beta$-energy distributions have an allowed shape. 
The reconstruction of the energy deposited in the event mimics that of the experiment. 
The experimental low energy threshold of 65~keV, 
is applied to each spectrometer segment before summing to obtain the total energy deposited.
Similarly a threshold (in both MC and experiment) of 105~keV is applied 
to the Si detector before gating on the spectrometer signals.

\begin{figure}[h]
\begin{center}
\includegraphics[width=8.6cm]{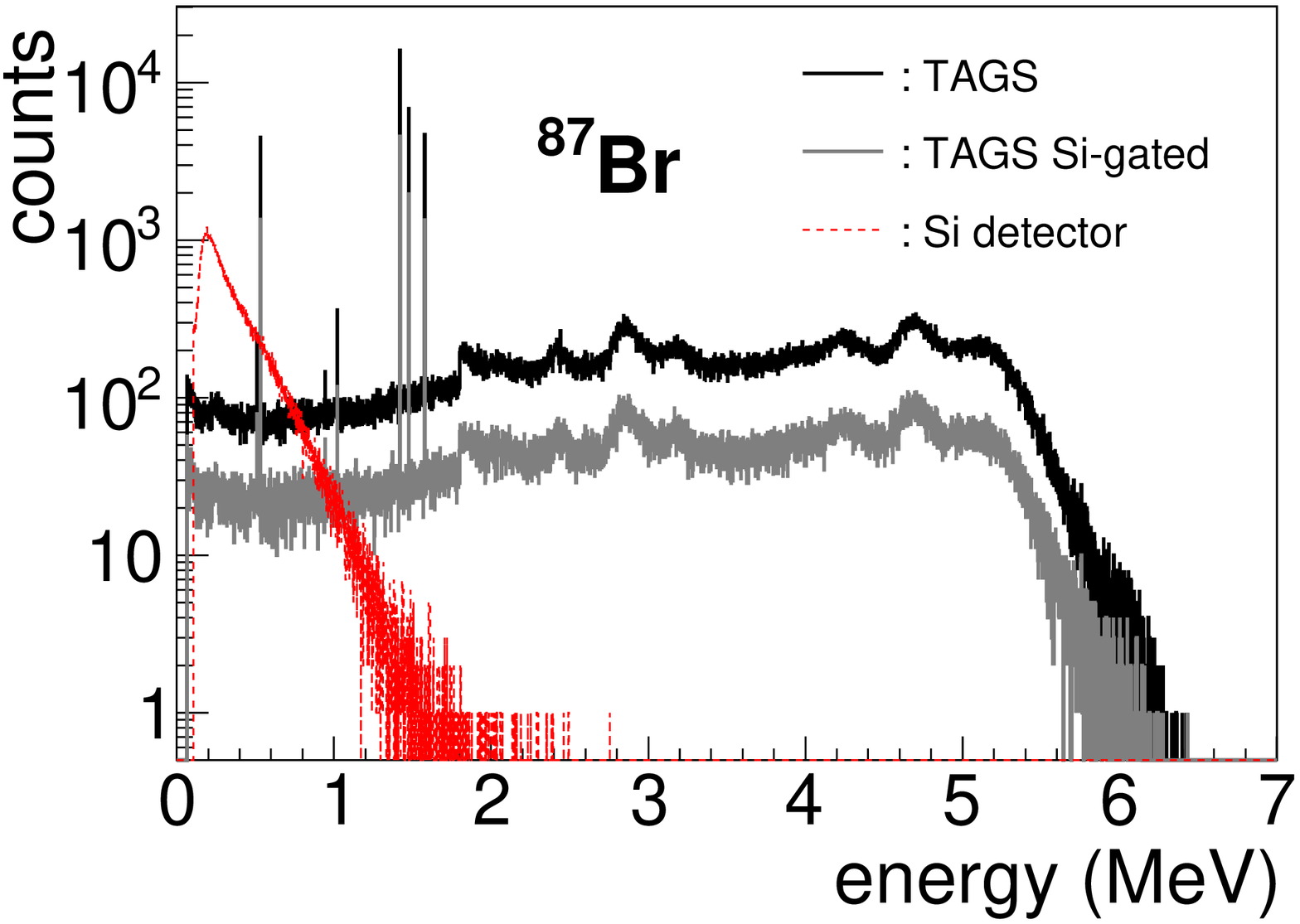}
\includegraphics[width=8.6cm]{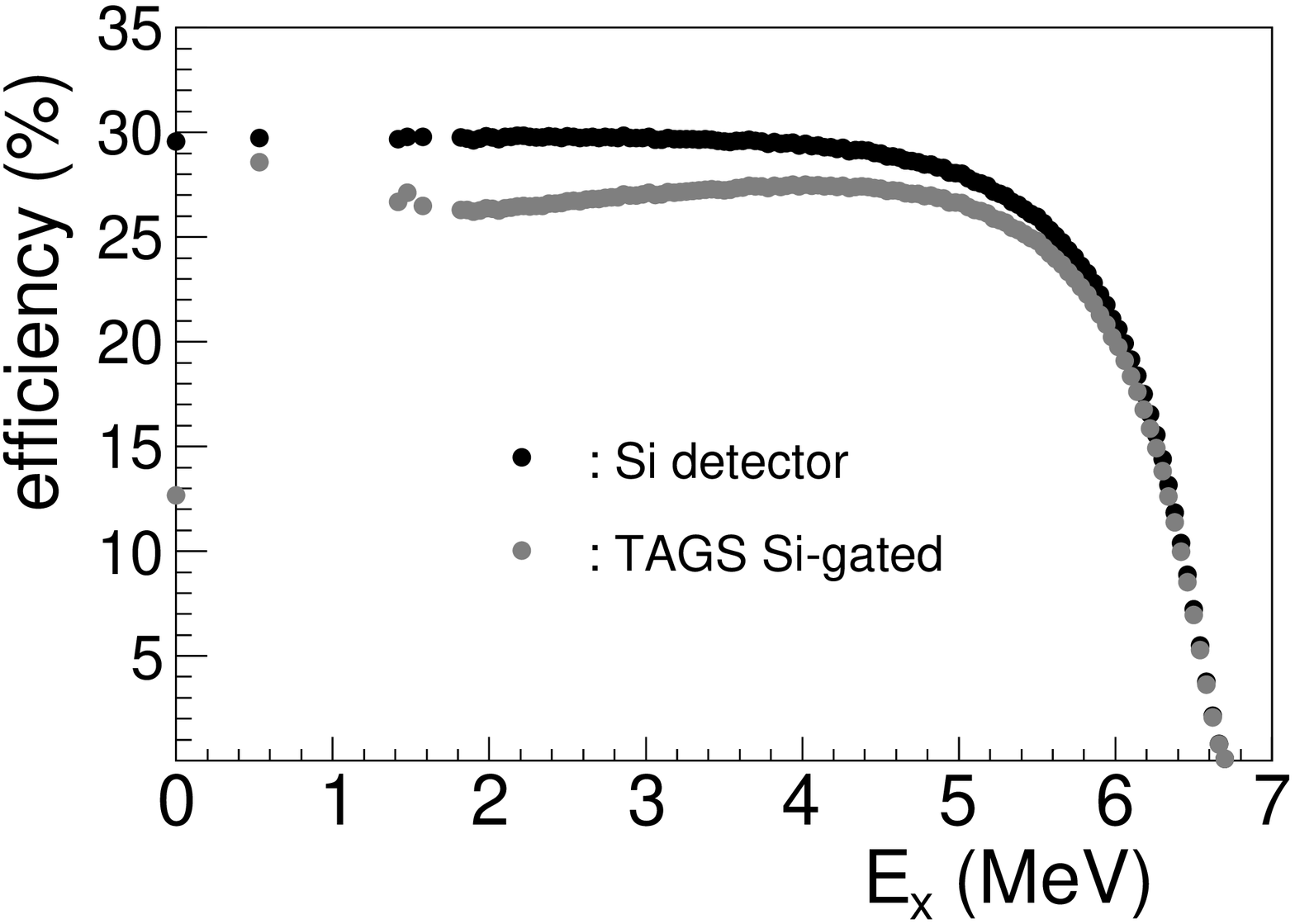}
\caption{(Color online)  Top panel: Simulated Si detector spectrum (red dashed line) and TAGS spectra with (gray line) 
and without (black line) gating on Si signals for the decay of $^{87}$Br. 
The instrumental resolution is not included in these spectra. Bottom panel: Simulated $\beta$
efficiency without (black symbols) and with (gray symbols) a gating condition on the $\gamma$
spectrometer. See text for further details.}
\label{fig1}
\end{center}
\end{figure}

In the top panel of Fig~\ref{fig1} the spectrum of energy deposited in the
spectrometer, with and without the $\beta$-gating condition, and the spectrum of energy
deposited in the Si detector are shown for the case of the decay of $^{87}$Br.
The instrumental resolution is switched-off in these spectra to show more clearly
their features.
One million decay events were simulated. The total number of counts
in the ungated TAGS spectrum is $0.890\times10^{6}$, and $0.255\times10^{6}$
in the gated spectrum. The counts in the Si detector spectrum are $0.288\times10^{6}$.
The lower panel of Fig.~\ref{fig1} shows the simulated detection efficiency of the
Si detector as a function of excitation energy together with the simulated probability
of registering a signal simultaneously in the Si detector and the $\gamma$ spectrometer.
The pronounced efficiency drop above $E_{x} = 5$~MeV is due to the low energy threshold
in the Si detector and the continuum nature of $\beta$ spectra.
The decrease of the spectrometer-gated Si detector efficiency
below 5~MeV in comparison with the ungated efficiency is due
to the importance in this decay of de-excitations proceeding by a single $\gamma$ transition
to the g.s., which have a greater probability of escaping detection in the spectrometer. 
Using these efficiency distributions and Eq.~\ref{eq2}
we can calculate the correction factors for each decay  
and apply Eq.~\ref{eq6} to obtain $I^{0}_{\beta}$.

The results of the calculation for the three isotopes are presented 
in Table~\ref{tabAp}. As can be observed the value obtained for the g.s.\ feeding
probability (10.12\%, 4.79\%, 0.11\% for $^{87}$Br, $^{88}$Br, $^{94}$Rb respectively) 
is very close to the true value 
(i.e. the input values for the simulation) in all cases (10.10\%, 4.72\%, 0\%).
If the original formula from Greenwood et al.\ (Eq.~\ref{eq5}) is used, 
with $N_{\beta \gamma}$ corrected for the $\beta$ penetration, we obtain 8.08\%, 3.69\% and 0.28\%, respectively, which deviate clearly from the true values
for the bromine isotopes.
From the values of the correction factor $c$ we can see that in this setup
the $\gamma$ spectrometer is rather sensitive to $\beta$ penetration
for decays to the g.s., between 48\% and 66\% of the average probability
of detecting a decay to an excited state. The increase in $c$ with isotope follows
the increase in $Q_{\beta}$.


\begin{table}
\begin{ruledtabular}
\begin{tabular}{cccccc} 
Isotope & R & a & b & c & $I^{0}_{\beta}$ [\%] \\  \hline
\rule{0pt}{3ex}
$^{87}$Br & 0.8846 & 1.064 & $4.91\times10^{-2}$ & 0.477 & 10.12 \\
\rule{0pt}{3ex}
$^{88}$Br & 0.9444 & 1.038 & $-1.67\times10^{-3}$ & 0.608 & 4.79 \\ 
\rule{0pt}{3ex}
$^{94}$Rb & 0.9913 & 1.008 & $-9.83\times10^{-3}$ & 0.656 & 0.11 \\ 
\end{tabular}
\caption{Ratio of counts $R$, correction factors $a$, $b$ and $c$ and 
calculated $\beta$ intensity to the
g.s for synthetic data. See text for further details.}
\label{tabAp}
\end{ruledtabular}
\end{table}


Another important check that can be performed with the synthetic data
is to test the stability of the result against uncertainties in the deconvolution procedure.
As explained in \cite{Val17} the $\beta$-intensity distribution obtained in the TAGS
analysis is affected by several systematic uncertainties related to the branching ratio matrix
used to build the spectrometer response, the accuracy of the MC simulations, the normalization of contaminants and even the deconvolution method.
This results in a spread of $I^{0}_{\beta}$ extracted from the deconvolution method
which varied between 9.16\% and 11.29\% (14 intensity distributions) for $^{87}$Br, 
and between 2.60\% and 5.82\% (13 intensity distributions) for $^{88}$Br. 
Actually this spread determines the size of the systematic uncertainty of the g.s.\ feed
obtained from the deconvolution method quoted above.
In comparison the statistical uncertainty from deconvolution is negligible (below 0.05\%).
However if we use the different $\beta$-intensity distributions to calculate the correction factors $a$, $b$ and $c$ and 
apply the $4\pi\gamma-\beta$ method to the synthetic spectra
simulated with the adopted $I_{\beta}$ distribution 
(shown in the top panel of
Fig.~\ref{fig1} for the $^{87}$Br case)
the resulting $I^{0}_{\beta}$ vary very little. 
For example in the deconvolution of $^{87}$Br data we
tested the effect of fixing the g.s.\ feeding probability to the ENSDF value 12\%.
This resulted in a still acceptable fit to the data, just outside the 5\%
maximum increase in $\chi^{2}$ that was used to select the set of acceptable
solutions in the original publication \cite{Val17}. 
When using the resulting
$\beta$ intensity to calculate the correction factors in Eq.~\ref{eq6} we obtain
$I^{0}_{\beta} = 10.17$\% very close to the true value. 
Likewise, in the case of $^{88}$Br we tested the effect 
in the deconvolution of fixing the g.s.\ feeding probability to 11\%, which is the upper
limit given by ENSDF. In this case the fit to the data was rather poor 
(44\% increase in $\chi^{2}$) as expected.
In spite of that, when we use the resulting $\beta$-intensity distribution 
to calculate the g.s.\ feeding probability in the $4\pi\gamma-\beta$ method
it gives 4.78\%, close to the  true value.
This simply reflects the fact that the TAGS technique determines rather accurately
the relative $\beta$ intensity to excited states that proceed by $\gamma$-ray emission,
on which the correction factors are based. 
In other words, the $4\pi\gamma-\beta$ method is much less sensitive
to the systematic uncertainties in the deconvolution of TAGS data.
An extreme example is presented in Section \ref{exper103}.

\bibliography{gsbib}

\end{document}